\documentclass[10pt]{article}
\usepackage[utf8]{inputenc}
\usepackage[T1]{fontenc}
\usepackage[english]{babel}
\usepackage{amsmath,amsthm,setspace, amsfonts, amssymb,graphics,multirow,enumerate}
\usepackage{todonotes}
\usepackage{graphicx,color}
\usepackage[linesnumbered,vlined,boxed,commentsnumbered]{algorithm2e}

 \usepackage{pdfpages}
\usepackage{float}
\usepackage{listings}
 \usepackage{bbm}
\usepackage{calc}
\usepackage{array}
 \usepackage{authblk}
 \usepackage{natbib}

\usepackage{graphicx}
\usepackage{pdfpages}
\usepackage{float}
\usepackage{listings}
 \usepackage{dsfont} 
 \usepackage{bbm}
\usepackage{amsmath}

\usepackage{calc}
\usepackage{array}
\usepackage{multirow}

\usepackage{natbib}

\usepackage{framed}

 \usepackage[pagewise]{lineno}

\begin{document}

\title{A SAEM Algorithm for Fused Lasso Penalized Non Linear Mixed Effect Models: Application to Group Comparison in pharmacokinetics}

\author[1,2,3]{Edouard Ollier}
\author[2]{Adeline Samson}
\author[3]{Xavier Delavenne}
\author[4]{Vivian Viallon}

\affil[1]{U.M.P.A., Ecole Normale Sup\'erieure de Lyon, CNRS UMR 5669; INRIA, Project-team NUMED. 46 All\'ee d'Italie, 69364 Lyon Cedex 07, France}
\affil[2]{Univ. Grenoble Alpes, LJK, F-38000 Grenoble, France; CNRS, LJK, F-38000 Grenoble; France}
\affil[3]{Groupe de Recherche sur la Thrombose, EA3065, Universit\'e de Saint-Etienne, Jean Monnet, F-42023, Saint-Etienne}
\affil[4]{Universit\'e de Lyon, F-69622, Lyon, France; Universit\'e Lyon 1, UMRESTTE, F-69373 Lyon; IFSTTAR, UMRESTTE, F-69675 Bron}

\maketitle

\begin{abstract}
Nonlinear mixed effect models are classical tools to analyze non linear longitudinal data in many fields such as population pharmacokinetics. Groups of observations are usually  compared by introducing the group affiliations as binary covariates with a reference group that is stated among the groups. This approach is relatively limited as it  allows only  the comparison of the reference group to the others. The proposed method compares groups  using a penalized likelihood approach. 
Groups are described by the same structural model but with parameters that are group specific. The likelihood is penalized with a fused lasso penalty that induces sparsity  in the differences between groups  for both fixed effects and variances of random effects. A penalized  Stochastic Approximation  EM algorithm is proposed that is coupled to Alternating Direction Method Multipliers to solve the maximization step. An extensive simulation study illustrates the performance of this algorithm when comparing more than two groups. Then the approach is applied to real data from two pharmacokinetic drug-drug interaction trials.
\end{abstract}

\section{Introduction}

Non Linear Mixed Effects Models (NLMEMs) are used to model and analyze longitudinal data in several fields, especially in clinical trials and population pharmacokinetic (PK). In clinical research, observations  may present a group structure  corresponding to the different treatment modalities. For example, a drug-drug interaction clinical trial between two compounds includes two groups of observations, patients treated with the molecule of interest and patients treated with the two compounds. The question is then to study the interaction between the two compounds. The example treated in this work is the analysis of data from two crossover trials studying the interaction between dabigatran etexilate (DE), an oral anticoagulant, and three P-glycoprotein inhibitors. The first trial is a  standard two way crossover trial with two treatment modalities: DE alone and DE plus a P-glycoprotein inhibitor. The second  one is an incomplete three way crossover trial with three treatment modalities:  DE alone, DE plus a first P-glycoprotein inhibitor and DE plus a second P-glycoprotein inhibitor. These two trials   study different dosing   regimens for DE and different P-glycoprotein inhibitors. Five groups of observations  can then be defined. The objective is to assess differences across these 5 groups corresponding to differences across the 5 treatment modalities. Usually, such a difference    is assessed through the variation of  the PK parameters across groups.  PK parameters are standardly estimated through an NLMEM. The difficulty, and this is the  objective of this paper, is then to identify   the significant differences between group's parameters. 

Statistical tests are classically used to identify significant influence  of the group structure on a PK parameter. The group affiliation is included as a categorical covariate and its influence is studied with maximum likelihood tests   \citep{samson2007saem, dubois2011model}.  Because the likelihood of NLMEM is intractable, stochastic versions of the EM algorithm such as the SAEM algorithm \citep{delyon1999convergence, kuhn2005maximum}, are generally used to estimate the model parameters. A stepwise procedure, based on the Bayesian Information Criterion (BIC), can then be performed to select the best model among the collection of models with the group affiliation covariate on each parameter. Two main drawbacks of this approach are that a reference group has first to be stated, and then only differences with regard to this reference group are considered. In the presence of more than two groups, this does not allow to select a model with no difference between two  ``non-reference'' groups. In order to study the differences between   non reference groups, combination of the group   covariates could be used, but their number increases rapidly with the number of groups. Indeed, the number of   between group differences models is equal to $(B_{G})^{p}$ where $B_{G}$ is Bell's number \citep{bell1934exponential} for $G$ groups and $p$ the number of studied parameters. Considering 5 groups and  studying between group differences on 3 parameters  leads to $52^{3}$ possible models.

Nevertheless, the relevance of group differences between all the groups can be directly studied using a penalized  joint modeling approach \citep{viallon2014robustness,oelker2014regularization,ollier2015regression}.  The same structural model is applied to each group with a structural sparsity-inducing penalty  \citep{bach2011convex} that encourages parameters to be similar in each group. These penalties are named \textit{structured penalties} and well-known examples are the group lasso \citep{bach2011convex} and the fused lasso  \citep{tibshirani2005sparsity}. In this work, we propose to estimate group parameters  by maximizing the penalized likelihood with a fused lasso penalty. This penalty was originally designed to penalize differences of coefficients corresponding to successive features   and has been generalized to account for features with a network structure \citep{hofling2010coordinate}.

Sparsity-inducing penalties in linear mixed effects models (LMEMs) have been proposed for selecting fixed effects only \citep{Schelldorfer2011, Rohart2014} and both fixed effects and random effects variances \citep{bondell2010joint}.  Note that  the joint selection of fixed effects and random effects variances is complex because the likelihood is  not convex with respect to the variances. The difficulty increases with NLMEMs as the likelihood is intractable (contrary to LMEMs), and only a few papers deal with penalties  in NLMEMs. \cite{arribas2014lasso} studied variable selection in semi parametric NLMEMs using a lasso penalty, the lasso selection step and the parameter estimation being realized separately. \cite{bertrand2015integrating} considered   $l_{1}$ penalized NLMEM for genetic variant selection. They proposed a penalized version of the SAEM algorithm,  in which the maximization step corresponds to an $l_{1}$ penalized  weighted least square problem. The recent stochastic proximal gradient algorithm \citep{atchade2014stochastic} has been applied to generalized LMEMs to optimize the likelihood penalized with a $l_{1}$ penalty. The  penalized likelihoods introduced in these papers are not based on a structured penalty in the sense that they do not  induce a structural sparsity. Up to our knowledge, no work investigates the use of a structured penalty in the context of NLMEMs.

The objective of this paper is to incorporate the fused lasso penalty in the SAEM algorithm, in particular to jointly estimate NLMEMs on several groups, and detect relevant differences among both fixed effects and variances of random effects. Penalties are introduced in the maximization step of the SAEM algorithm. Fixed effects and variances of random effects are penalized through a sum of absolute differences. The penalized differences correspond to edges of a graph in which the vertices correspond to the groups. Solving this penalized optimization problem is not trivial and we suggest to use an Alternating Direction Method of Multipliers (ADMM) \citep{boyd2011distributed}.  The direct penalization of the variances leading to a non convex optimization problem, we propose to penalize the inverse  of the covariance matrix, assuming  this matrix is diagonal. An ADMM algorithm can be used to solve the corresponding penalized optimization problem, its proximal step being explicit or not, depending on the number of groups. We also consider weighted penalties, following the ideas of the adaptive Lasso \citep{zou2006adaptive}.  Selection of the two tuning parameters introduced in the two penalties is   performed according to the BIC.  

The paper is organized as follows. Section \ref{sec:JOINT} introduces   NLMEM and the SAEM algorithm.  In Section \ref{sec:PEN} we introduce the fused lasso penalty, the penalized SAEM algorithm.  The tuning parameter selection is described in Section \ref{sec:BIC}. In Section \ref{sec:SIM}, the penalized-SAEM algorithm  is  evaluated on simulated data  with 2 groups or more. Finally, it is applied   on real data from the crossover clinical   trials studying drug-drug interaction between dabigatran etexilate and   three other drugs in Section  \ref{sec:CROSS}.

\section{Joint estimation of multiple nonlinear mixed effects models}
\label{sec:JOINT}
 
Let $y_{g,i,j}$ be the observation at time $t_{g,i,j}$ ($j \in \{1,\ldots,n_{g,i}\}$) for the $i$-th patient ($i \in \{1,\ldots,N_{g}\}$) in the $g$-th group ($g \in \{1,\ldots,G\}$). We consider models of the form:
\begin{eqnarray}
&& y_{g,i,j} = f(t_{g,i,j},\phi_{g,i}) +  d(t_{g,i,j},\phi_{g,i}) \varepsilon_{g,i,j} \nonumber \\
&& \varepsilon_{g,i,j} \sim \mathcal{N}(0,1) \mbox{ (iid), } \nonumber 
\end{eqnarray}
where $f$ and $d$ are two given nonlinear functions. The function $d$ corresponds to the error model. In this paper, we restrict to the standard form  $d=af + b$ with $a$ and $b$ two constants. Measurement errors $\varepsilon_{g,i,j}$ are further assumed to be independent and identically distributed. Individual parameters $\phi_{g,i}$ for the $i$-th subject in group $g$ is a $p$-dimensional random vector, independent of  $\varepsilon_{g,i,j}$ and  assumed to be decomposable (up to a transformation $h$) as:
\begin{eqnarray}
h(\phi_{g,i}) &=& \mu_{g}  + b_{g,i} \nonumber \\
b_{g,i} &\sim& \mathcal{N}(0,\Omega_{g}) \mbox{ (iid)}. \nonumber
\end{eqnarray}
Here $\mu_{g} \in \mathbbm{R}^{p}$ is the mean parameter  vector for  group $g$ and $b_{g,i} \in \mathbbm{R}^{p} $ the random effects of the $i$-th patient.  Various transformations $h$ can be used. Here we use the common one $h(x) = \log(x)$,  which yields log-normally distributed $\phi_{g,i}$. In this work,  $\Omega_{g}$ is supposed diagonal as explained in section \ref{sec:PEN}.

The log-likelihood then takes the form:
\begin{linenomath}
 \begin{gather}\label{eq:likelihood}
LL(\theta) =\log  p(y;\theta)= \sum_{g=1}^{G}  \log \left(  \int p(y_{g},\phi_{g} ; \theta_{g}) d\phi_{g} \right),
\end{gather}
where $p(y_{g},\phi_{g} ; \theta_{g})$ is  the likelihood of the complete data in group $g$:
 \begin{gather*}
\log p(y_{g},\phi_{g} ; \theta_{g}) =   -\sum_{i,j} \log(d(t_{g,i,j},\phi_{g,i}) ) - \frac{1}{2}\sum_{i,j} \left( \frac{y_{ij} - f(t_{g,i,j},\phi_{g,i}) }{d(t_{g,i,j},\phi_{g,i})} \right)^{2}  -\frac{N_{g}}{2}\log(\vert \Omega_{g} \vert) \nonumber \\ - \frac{1}{2} \sum_{i} (\phi_{g,i} - \mu_{g})^{t}\Omega_g^{-1}(\phi_{g,i} - \mu_{g}) - \frac{\sum_{i} n_{g,i} + N_{g}p}{2}\log(2\pi), \nonumber
\end{gather*}
\end{linenomath}
with $\theta = (\theta_{1},\ldots,\theta_{G})$ and $\theta_{g} = (\mu_{g},\Omega_{g},a,b)$ the parameters to be estimated. Note that the log-likelihood $LL(\theta)$ as defined in Equation (\ref{eq:likelihood}) has generally no closed form expression because of the  nonlinearity with respect to $b_{g,i}$.  


 In this section, we present a standard version of the SAEM algorithm in the context of joint estimation,  on which the penalized version that we introduce later will be based.  For now, we do not account for potential similarities of the parameters across groups. The SAEM algorithm is a classical tool for parameter  estimation of NLMEMs  \citep{delyon1999convergence}. It iteratively maximizes the conditional expectation of the complete data log-likelihood. At iteration $k$, and given the current estimate $\theta_{k-1}$, the problem reduces to the optimization of the following criterion:
 \begin{linenomath}
 \begin{gather*}
Q_{k}(\theta) = \sum_{g=1}^{G}  Q_{g,k}(\theta_{g}) =  \sum_{g=1}^{G} \mathbb{E} \left(\; \log p(y_{g},\phi_{g};\theta_{g}) \; \vert \; y_{g}, \theta_{g,k-1} \right). \nonumber
\end{gather*}
\end{linenomath}
As this conditional expectation has no closed form for NLMEMs, it is approximated using a stochastic approximation scheme. The E-step of the classical EM algorithm is then divided in two parts: a simulation step where individual parameters are simulated using a Markov Chain Monte Carlo method (MCMC), and a stochastic approximation step  \citep{kuhn2005maximum}. At   iteration $k$ of the SAEM algorithm we have:
\begin{enumerate}
\item Estimation step (E-step):
\begin{enumerate}
\item Simulation step: draw $\phi_{g,k}$   using an MCMC procedure  targeting $p(. \vert y_{g} , \theta_{g,k-1})$.
\item Stochastic approximation    step of $Q_{k}(\theta)$: update  $\tilde{Q}_{k}(\theta)$ using the following scheme
\begin{linenomath}
 \begin{gather*}
\tilde{Q}_{g,k}(\theta_{g}) = \tilde{Q}_{g,k-1}(\theta_{g}) + \gamma_{k}(\log\mbox{ }p(y_{g},\phi_{g,k};\theta_{g}) - \tilde{Q}_{g,k-1}(\theta_{g}) ),  \nonumber
\end{gather*}
\end{linenomath}
where $\gamma_{k}$ is a decreasing sequence of positive numbers. When the complete data likelihood belongs to the exponential family, this step simply reduces to the stochastic approximation of its sufficient statistics $s_{g,1,k}$, $s_{g,2,k}$ and  $s_{g,3,k}$:
\begin{linenomath}
\begin{eqnarray}
s_{g,1,k} &=& s_{g,1,k-1} + \gamma_{k}\left( \sum_{i=1}^{N_{g}}\phi_{g,i,k} - s_{g,1,k-1} \right) \nonumber \\
s_{g,2,k} &=& s_{g,2,k-1} + \gamma_{k} \left( \sum_{i=1}^{N_{g}} \phi_{g,i,k} \phi_{g,i,k}^t - s_{g,2,k-1} \right)\nonumber \\
s_{g,3,k} &=&\begin{cases} s_{g,3,k-1} + \gamma_{k} \left( \sum_{i,j} \left( y_{g,i,j} - f(t_{g,i,j},\phi_{g,i,k}) \right)^{2}  - s_{g,3,k-1} \right) \mbox{ if } b=0 \\  s_{g,3,k-1} + \gamma_{k} \left( \sum_{i,j} \left( \frac{y_{g,i,j} - f(t_{g,i,j},\phi_{g,i,k})}{d(t_{g,i,j},\phi_{g,i,k})} \right)^{2}  - s_{g,3,k-1} \right) \mbox{ if } a=0 \end{cases}.\nonumber 
\end{eqnarray}
\end{linenomath}
\end{enumerate}  
\item Maximisation step (M-step): update of population parameters: 
 \begin{eqnarray}
\theta_{k} =   \underset{\theta} {\operatorname{ArgMax}} \mbox{ } \tilde{Q}_{k}(\theta). \nonumber
\end{eqnarray}
Within the exponential family,   the    solution is explicit  for $\mu_{g}$ and $\Omega_{g}$:
\begin{eqnarray}
&&\mu_{g,k} =\frac{1}{N_{g}}s_{g,1,k}\; \mbox{ and }  
\Omega_{g,k} = \frac{1}{N_{g}}  \left( s_{g}^{2,k}   -   \sum_{i=1}^{N_{g}}  \mu^{g}_{k}  s_{gi}^{1,k ^{t}} - \sum_{i=1}^{N_{g}}  s_{gi}^{1,k}  \mu^{t}_{gk} \right)  +  \mu_{gk}  \mu^{t}_{gk}. \nonumber
\end{eqnarray}
For   parameters $a$ and $b$,  they are updated using the whole data set because they are common to all  groups. An explicit solution exists when $a=0$ or $b=0$:
\begin{eqnarray}
&& a = 0 \Rightarrow b_{k} =  \sqrt{\frac{\sum_{g=1}^{G}s_{g,3,k}}{\sum_{g=1}^{G} \sum_{i} n_{g,i} }} \nonumber \\
&& b = 0 \Rightarrow a_{k} = \sqrt{\frac{\sum_{g=1}^{G}s_{g,3,k}}{\sum_{g=1}^{G} \sum_{i} n_{g,i} }}.  \nonumber
\end{eqnarray}
When $a\neq0$ and $b\neq0$, the maximization problem has to be solved numerically.
\end{enumerate}  
Thus, except for  $a$ and $b$,  the SAEM algorithm for the joint estimation problem is implemented as if the  $G$ groups were analyzed separately.

\section{Penalized joint estimation of  group-structured NLMEM}
\label{sec:PEN}

The previous SAEM algorithm   corresponds to  parameters estimated within each group. But  groups can be expected to share common characteristics, so that theoretical parameters are expected to exhibit similarities. Therefore, we introduce a penalty within the SAEM algorithm that encourages parameters to be equal. We detail the fused penalties and the penalized SAEM algorithm.

\begin{figure}
\begin{center}
\includegraphics[scale = 0.75]{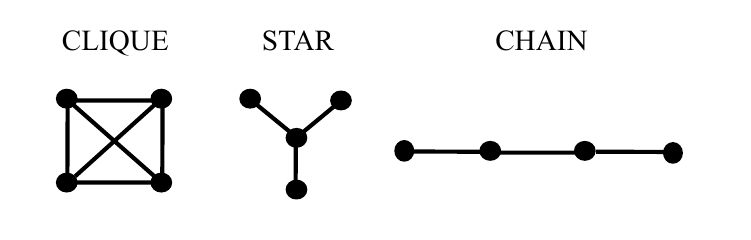}
\caption{Examples of graphs for $G=4$ groups}
\label{fig:GraphJoint}
\end{center}
\end{figure}

The fused lasso penalty   encourages parameters to have the same value between two  groups. This  is   particularly useful when theoretical parameters of (at least some of) the groups are expected to be similar and/or when the objective of the study is to assess potential differences between groups.  Depending on the context, differences between all the groups or only some specific differences might be of interest. Likewise,  similarity of some parameters does not necessarily hold for all the groups. 
These differences and similarities of interest can be described with a graph that links groups together. Two groups are related in the graph if the comparison of these two groups is of interest, or if parameters are assumed to be similar in these two groups. Of course, any graph structure can be put forward, but some of them are naturally appealing in various contexts (see figure \ref{fig:GraphJoint} with $G=4$): 
\begin{itemize}
\item Clique Graph: no  assumption on the hierarchical structure of the groups are made. All the possible differences between group parameters are penalized.
\item Star Graph:   a reference group is stated and only the differences between the reference group and the others are penalized. This   is equivalent to the  standard approach based on group affiliation covariate.
\item Chain Graph: when groups can  naturally be ordered.
\end{itemize}
Given a specific graph described by its edge set $ \mathcal{E}$, we introduce the penalties for the fixed and the variance parameters. 

 For fixed parameters $(\mu_{1},\ldots,\mu_{G})$, the fused lasso penalty corresponds to:
 \begin{eqnarray}
&&P_{F}(\mu_{1},\ldots,\mu_{G}) =  \sum_{(g_{1},g_{2})\in \mathcal{E}}  \Vert \mu_{g_{1}} - \mu_{g_{2}} \Vert_{1},  \nonumber 
 \end{eqnarray}
where $\Vert x \Vert_{1}=\sum_{i} \vert x_{i} \vert$ is the $l_{1}$-norm. The fused lasso penalty encourages  the fixed parameters of two groups connected in the graph to be equal.

Concerning random effect variances,  a natural idea  would be to penalize  them directly. However, the resulting  optimization problem is not convex, as the Gaussian  complete likelihood is the sum of a concave and a convex function.  This makes this problem intractable with standard tools from convex analysis. Some algorithms have been proposed to solve simple $l_{1}$-penalized problems \citep{bien2011sparse,wang2013coordinate} but their extension to the fused penalty context is not straightforward and they are computationally demanding. As the solver has to be called within each iteration of SAEM, this would lead to an unreasonable computing time. However under the assumptions that $\Omega_{g}$ is diagonal for each group $g$, we have $\Omega_{g_1}=\Omega_{g_2}$ if and only if $\Omega^{-1}_{g_1} = \Omega^{-1}_{g_2}$. Then a simple alternative consists in penalizing the inverse of the covariance matrix, that is the precision matrix. Indeed the  corresponding penalized optimization problem becomes convex and can be solved efficiently. On the other hand, when matrix $\Omega_{g}$ is not diagonal for all $g$, the support of the differences among variances has no guarantee to be equal to the support of the differences among the diagonal elements of the precision matrices. Therefore we focus on the diagonal case here and we use the following penalty:
\begin{linenomath}
\begin{gather*}
P_{V}(\Omega_{1}^{-1},\ldots\Omega_{G}^{-1})  =  \sum_{(g_{1},g_{2})\in \mathcal{E}} \Vert \Omega_{g_{1}}^{-1} -\Omega_{g_{2}}^{-1} \Vert_{1} = \sum_{i=1}^{p} \sum_{(g_{1},g_{2})\in \mathcal{E}} \vert \Omega_{g_{1},ii}^{-1} -\Omega_{g_{2},ii}^{-1} \vert. \nonumber
 \end{gather*}
 \end{linenomath}
Of course, penalizing differences between $\Omega_g^{-1}$ is not equivalent to penalizing differences between $\Omega_{g}$ as $\vert \Omega_{g_{1},ii}^{-1} -\Omega_{g_{2},ii}^{-1} \vert \neq \vert \Omega_{ii}^{g_{1}} -\Omega_{ii}^{g_{2}} \vert $. Some issues may occur when considering parameters with very different levels of variability, our proposal being more likely to discard differences for parameters with low variances. This issue is mitigated when working with log-normally distributed individual parameters.  Adaptive weights can further help to prevent such a behavior (see section \ref{sec:SIM}).
 
Weights ($\pi$,$\nu$) can be introduced in order to account for potential prior information:
\begin{linenomath}
 \begin{eqnarray}
&&P_{F}(\mu_{1},\ldots,\mu_{G}) =  \sum_{(g_{1},g_{2})\in \mathcal{E}} \sum_{i=1}^{p} \pi_{g_{1}g_{2},i}  \vert \mu_{g_{1},i} - \mu_{g_{2},i}  \vert  \nonumber \\
&&P_{V}(\Omega_{1}^{-1},\ldots\Omega_{G}^{-1}) =  \sum_{(g_{1},g_{2})\in \mathcal{E}} \sum_{i=1}^{p} \nu_{g_{1}g_{2},i}  \vert \Omega_{g_{1},ii}^{-1} -\Omega_{g_{2},ii}^{-1} \vert. \nonumber
 \end{eqnarray}
 \end{linenomath}
These weights can be based on initial maximum likelihood estimates within each group ($\tilde{\mu}_{g}$,$\tilde{\Omega}_{g}$) following the idea of the adaptive fused lasso \citep{viallon2014robustness}:  $\pi_{g_{1}g_{2},i} = \vert \tilde{\mu}_{g_{1},i} - \tilde{\mu}_{g_{2},i}  \vert^{-\alpha}$ and $\nu_{g_{1}g_{2},i} = \vert \tilde{\Omega}_{g_{1},ii}^{-1} - \tilde{\Omega}_{g_{2},ii}^{-1} \vert^{-\alpha}$ for some $\alpha > 0$ (typically  $\alpha=1$).
These weighted penalties are particularly helpful to compute unpenalized re-estimation of  the selected model (see Section \ref{sec:BIC}). Finally, observe that these  weighted penalties with weights $\pi$ and $\nu$ can be written in a matrix form:
 \begin{linenomath}
 \begin{gather*}
P_{F}(\mu_{1},\ldots,\mu_{G})   = \Vert \pi \circ P  \mu  \Vert_{1} \nonumber \\
P_{V}(\Omega_{1}^{-1},\ldots\Omega_{G}^{-1})   = \Vert \nu \circ P  \mbox{diag}(\Omega^{-1})  \Vert_{1}, \nonumber 
 \end{gather*}
 \end{linenomath}
 where the matrix $P\in \{-1,0,1\}^{\vert \mathcal{E} \vert \times Gp}$ encodes the penalized values of $\mu = (\mu_{1},\ldots,\mu_{G})^{t}$ and $\mbox{diag}(\Omega^{-1}) = (\mbox{diag}(\Omega_{1}^{-1}),\ldots,\mbox{diag}(\Omega_{G}^{-1}))^{t}$   and $\circ$ stands for the Hadamard product.

The penalized   SAEM algorithm consists in iteratively maximizing the penalized stochastic approximation of the conditional expectation $Q_{k}(\theta)$:
\begin{linenomath}
 \begin{gather}
\tilde{Q}_{k}(\theta) - \lambda_{F} P_{F}(\mu_{1},\ldots,\mu_{G}) -  \lambda_{V}  P_{V}(\Omega_{1}^{-1},\ldots\Omega_{G}^{-1}), \nonumber
\end{gather}
\end{linenomath}
where $\lambda_{F}$ and  $\lambda_{V}$ are two tuning parameters to be calibrated (see Section \ref{sec:BIC}) that control the penalty strength. When  $\lambda_{F}=\lambda_{V}=0$, the estimates correspond to the standard maximum likelihood estimates. For large enough values, the vector of penalized differences is set to zero ($P  \mu =0$ and$/$or $P  \mbox{diag}(\Omega^{-1})  =0$). 

The penalized SAEM is the standard SAEM except for the M-step: 
 a fused lasso penalized regression problem is solved for both fixed effects and random effects variances updates, with fixed tuning parameters $\lambda_{F}$ and  $\lambda_{V}$. At iteration $k$, it corresponds to (Box $1$): 
\begin{figure}[H]
\begin{framed}
{\bf Box $1$: Maximization step of the penalized SAEM algorithm}
\begin{enumerate}
\item Fixed effects update:
$$\left( \mu_{1,k},\ldots,\mu_{G,k} \right) =  \underset{\mu_{1},\ldots,\mu_{G}} {\operatorname{ArgMax}} \left(\sum_{g=1}^{G} \tilde{Q}_{g,k}(\mu_{g},\Omega_{g,k-1},a_{k-1},b_{k-1}) \; -  \;  \lambda_{F}P_{F}(\mu_{1},\ldots,\mu_{G}) \right). $$
\item  Random effects variances update:
$$(\Omega^{1}_{k},\ldots,\Omega^{G}_{k}) =  \underset{\Omega^{1},\ldots\Omega^{G}} {\operatorname{ArgMax}}\left(\sum_{g=1}^{G} \tilde{Q}_{g,k}(\mu_{g,k}, \Omega_{g},a_{k-1},b_{k-1})   \; -  \;  \lambda_{V}P_{V}(\Omega_{1}^{-1},\ldots, \Omega_{G}^{-1}) \right).$$
\item   Error model parameters update:   usual update.
\end{enumerate}
\end{framed}
\end{figure}
We now turn to the description of the two update steps, for the fixed effects and  the random effects variances respectively.


For fixed effects update, the conditional expectation of the complete likelihood reduces to the following weighted least square function:
\begin{linenomath}
$$\tilde{Q}_{k}(\mu) = \sum_{g=1}^G \tilde{Q}_{k}(\mu_{g},\Omega_{g,k-1},a_{k-1},b_{k-1}) =  C -\frac{1}{2} \sum_{g=1}^G  \left(   -  \mu_{g}^{t}  \Omega_{g,k-1}^{-1} s_{g,1,k}  - s_{g,1,k}^{t} \Omega_{g,k-1}^{-1}  \mu_{g}  +  \mu_{g}^{t}  \Omega_{g,k-1}^{-1}\mu_{g} \right),  $$
\end{linenomath}
where $C$ is a constant not depending on $\mu_{g}$. The matrix form of the problem to be solved is:
\begin{linenomath}
\begin{eqnarray}
\left( \mu_{1,k} ,\ldots, \mu_{G,k}  \right) 
&=&   \underset{ \mu } {\operatorname{ArgMax}} \mbox{ } \left( \tilde{Q}_{k}(\mu)  -  \lambda_{F} \Vert P \mu  \Vert_{1} \right). \label{eq:mupenalized}
\end{eqnarray}
\end{linenomath}
This optimization problem corresponds to an extension of the generalized fused lasso of  \cite{hofling2010coordinate} with least squares replaced by weighted least squares. It can be solved with the Alternating Direction Method of Multipliers (ADMM) \citep{boyd2011distributed}, that breaks the convex optimization problem   into small pieces. We briefly recall the idea of ADMM using the standard  ADMM notations. Problem (\ref{eq:mupenalized}) can be rewritten as an equality constraints optimization problem, where $\mu$ is   split in two parts $\mu$ and $z$:
\begin{linenomath}
\begin{eqnarray}
&& \hat{\mu} =  \underset{ \mu,z } {\operatorname{ArgMin}} \mbox{ } \left(   -\tilde{Q}_{k}(\mu)  + \lambda_{F} \Vert z \Vert_{1} \right). \nonumber \\
&& \mbox{ s.t } P \mu  - z = 0 \nonumber
\end{eqnarray}
\end{linenomath}
The ADMM algorithm solves  (\ref{eq:mupenalized}) by iteratively solving smaller (and easier) problems for each primal ($\mu$, $z$) and dual ($u$) variables separately using the   augmented Lagrangian formulation:
\begin{linenomath}
\begin{gather*}
\underset{ \mu,z } {\operatorname{ArgMin}} \mbox{ }  \underset{ u } {\operatorname{ArgMax}} \mbox{ } \left(  -\tilde{Q}_{k}(\mu)  + \lambda_{F} \Vert z \Vert_{1}  +  \langle u, P \mu - z  \rangle + \frac{\rho}{2} \Vert P \mu - z + u \Vert^{2}_{2} \right). \nonumber
\end{gather*}
\end{linenomath}
Here $\rho$ is the augmented Lagrangian parameter (generally set to $1$) and $\Vert \cdot \Vert_{2}$ the $l_{2}$-norm. The ADMM algorithm  consists in applying the steps  presented in Box $2$ at each iteration $q$ until convergence. 
\begin{figure}
\begin{framed}
{\bf Box $2$: ADMM algorithm for fixed effects update }
\begin{enumerate}
\item Initialization: $\mu_{0}=\mu_{k-1}$,  $z_{0}=0$, $u_{0}=0$ 
\item For $q=0,1,2,...$ until convergence:
\begin{enumerate}
\item $\mu$ update:
$$ \mu_{q+1} = \underset{ \mu} {\operatorname{ArgMin}} \left( -\tilde{Q}_{k}(\mu)  + \frac{\rho}{2} \Vert P \mu - z_{q} + u_{q} \Vert^{2}_{2} \right) = (  \Delta  + \rho P^{t}P )^{-1}(\Gamma + \rho P^{t}(z_{q} - u_{q})), 
$$
with $\Gamma =  \mbox{diag}(  \Omega_{1,k-1}^{-1} s_{1,1,k},\ldots,   \Omega_{G,k-1}^{-1} s_{G,1,k})$ \\ and $\Delta =\mbox{diag}(N_{1} \Omega_{1,k-1}^{-1},\ldots,N_{G} \Omega_{G,k-1}^{-1})$
\item $z$ update:
$$ z_{q+1}  = \underset{ z} {\operatorname{ArgMin}} \left(   \frac{\rho}{2} \Vert P  \mu_{q+1} + u_{q} - z  \Vert^{2}_{2}  + \lambda_{F} \Vert z \Vert_{1} \right) =  \mathcal{S}_{\frac{\lambda_{F}}{\rho}} (P \mu_{q+1} + u_{q} ),
$$
with the soft thresholding operator   $\mathcal{S}_{\lambda} (x) = sgn(x)(\vert x \vert - \lambda)_{+}$.
\item dual update:
 \begin{eqnarray} u_{q+1} = u_{q} + P \mu_{q+1} - z_{q+1}. \nonumber 
  \end{eqnarray}
\end{enumerate} 
\end{enumerate} 
\end{framed}
\end{figure}
When adaptive weights are included in the penalty, the same algorithm can be used except that the tuning parameter $\lambda_{F}$ is replaced by the vector $\lambda_{F} \circ \pi$.
$ $\\

Concerning random effects covariance matrix update, the conditional expectation of the complete likelihood for group $g$ is:
\begin{linenomath}
\begin{eqnarray}
\tilde{Q}_{k}(\mu_{g,k},\Omega_{g},a_{k-1},b_{k-1})  = C - \log \vert \Omega_{g} \vert - \mbox{Trace} \left[ \Omega_g^{-1} \tilde{\Sigma}_{g,k}  \right], \nonumber
\end{eqnarray}
\end{linenomath}
where $C$ is a constant not depending on $\Omega_{g}$, and $\tilde{\Sigma}_{g,k}=   \frac{1}{N_{g}}  \left( s_{g}^{2,k}   -   \sum_{i=1}^{N_{g}}  \mu^{g}_{k}  s_{gi}^{1,k ^{t}} - \sum_{i=1}^{N_{g}}  s_{gi}^{1,k}  \mu^{t}_{gk} \right)  +  \mu_{gk}  \mu^{t}_{gk}$ corresponds to the solution of the unpenalized problem. Then the problem to be solved is:
\begin{linenomath}
\begin{gather}
(\Omega^{1}_{k},\ldots\Omega^{G}_{k}) =  \underset{\Omega^{1},\ldots\Omega^{G}} {\operatorname{ArgMax}}\left(-\sum_{g=1}^{G}  \left(\log \vert \Omega_{g} \vert + \mbox{Trace}  \left[ \Omega_g^{-1} \tilde{\Sigma}_{g,k}  \right] \right) \; -\;   \lambda_{V}P_{V}(\Omega_{1}^{-1},\ldots, \Omega_{G}^{-1}) \right). \label{eq:omegapenalized}
\end{gather}
\end{linenomath}  
\cite{danaher2011joint} consider a similar optimization problem (for joint graphical models) and propose an ADMM algorithm to solve it. We apply the same methodology here. We briefly recall its principle here and refer to \cite{danaher2011joint} for more details. Problem (\ref{eq:omegapenalized}) has the following  scaled augmented Lagrangian: 
\begin{linenomath}
\begin{gather*}
\underset{  \Omega_g^{-1}, \, Z_{g}} {\operatorname{ArgMin}} \mbox{ }  \underset{ U_{g} } {\operatorname{ArgMax}} \mbox{ }  \begin{cases}   \sum_{g=1}^{G}  \left(\log \vert \Omega_{g} \vert + \mbox{Trace} \left[ \Omega_g^{-1} \tilde{\Sigma}_{g,k}  \right] \right)  + \lambda_{V}P_{V}(Z_{1},\ldots, Z_{G}) \\  + \sum_{g=1}^{G}  \frac{\rho}{2} \Vert \Omega_g^{-1}  - Z_{g} + U_{g} \Vert^{2}_{F} \end{cases},  \nonumber
\end{gather*}
\end{linenomath}
where  $(\Omega_g^{-1})_{g=1,...,G}$,$(Z_{g})_{g=1,...,G}$ are the primal variables,  $(U_{g})_{g=1,...,G}$ the dual variables, $\rho$ is the augmented Lagrangian parameter (generally set to $1$) and $\Vert X \Vert_{F}^{2}$ corresponds to the Frobenius norm of  matrix $X$. The ADMM algorithm  consists in applying the steps  presented in Box $3$ at each iteration $q$ until convergence.
\begin{figure}
\begin{framed}
{\bf Box $3$: ADMM algorithm for variances update}
\begin{enumerate}
\item Initialization: $\Omega_{g,0}=\tilde{\Sigma}_{g,k}$,  $Z_{g,0}=0$, $U_{g,0}=0$ 
\item For $q=0,1,2,...$ until convergence:
\begin{enumerate}
\item $\Omega$ update: for $g=1,\ldots,G$
\begin{gather*}
\Omega^{-1}_{g,q+1} = \underset{\Omega_g^{-1}} {\operatorname{ArgMin}} \mbox{ }  \left(\log \vert \Omega_{g} \vert + \mbox{Trace}(\tilde{\Sigma}_{g,k} \Omega_g^{-1}) + \frac{\rho}{2} \Vert \Omega_g^{-1} - Z_{q} + U_{q} \Vert^{2}_{F}  \right).
\end{gather*}
\item $Z$ update: 
\begin{gather*}
(Z_{1,q+1},\ldots,Z_{G,q+1}) =  \underset{ Z} {\operatorname{ArgMin}} \left(   \sum_{g=1}^{G}  \frac{\rho}{2} \Vert \Omega^{-1}_{g,q+1}  - Z_{g} + U_{g,q} \Vert^{2}_{F}   + \lambda_{V}P_{V}(Z_{1},\ldots, Z_{G}) \right). \end{gather*}
\item  dual update: for $g=1,\ldots,G$
\begin{gather*}
U_{g,q+1} = U_{g,q} + \Omega^{-1}_{g,q+1}  - Z_{g,q+1}.
\end{gather*}
\end{enumerate}
\end{enumerate}
\end{framed}
\end{figure}
Step 2(a)  has an explicit solution \citep{witten2009covariance}. Step  2(b) is the evaluation of the $P_{V}$'s proximal operator. An explicit formula is available when $G=2$ \citep{danaher2011joint}, but for $G>2$ it has to be numerically approximated. This   increases computational time significantly. As for fixed effects, when adaptive weights are included in the penalty, the same algorithm can be used except that the tuning parameter $\lambda_{V}$ is replaced by the vector $\lambda_{V} \circ \nu$.

\section{Selection of the tuning parameters and final estimator}
\label{sec:BIC}

The described SAEM algorithm is applied with a fixed value of the tuning parameters  $\Lambda = (\lambda_{F}, \lambda_{V})$.  The value of these tuning parameters varying from zero to infinity, the SAEM algorithm selects a collection of models with a  typically  decreasing number of  between-group differences (from the full model to the model with no difference at all). The optimal $\Lambda$ can be selected using   the Bayesian Information Criterion  (BIC):   the optimal $\Lambda$  is defined as the one corresponding to the model with minimal BIC. In practice, we first perform a manual search in order to determine $(\lambda^{MAX}_{F}, \lambda^{MAX}_{V})$, which are the smallest values for which all the penalized differences are null. Then we  run the algorithm on a user-defined grid $(\Lambda_{1},\ldots, \Lambda_{M}) \in ([ 0, \lambda^{MAX}_{F} ] \times  [ 0, \lambda^{MAX}_{V} ])^{M}$. The optimal value  $\Lambda_{BIC}$ is defined as:  
\begin{linenomath}
\begin{gather*}
\Lambda_{BIC} =  \underset{ \Lambda \in \{\Lambda_{1},\ldots, \Lambda_{M} \} } {\operatorname{ArgMin}} \mbox{ } BIC(\Lambda), \nonumber
\end{gather*}
\end{linenomath}
where $BIC(\Lambda)$ is the criterion of the model corresponding to the value $\Lambda$. For a NLMEM  with random effects on all the parameters,  the BIC is generally defined as   \citep{delattre2014note}:
\begin{linenomath}
\begin{gather*}
  BIC = -2LL(\theta) + \log(N)\times df(\theta),  \nonumber
\end{gather*}
\end{linenomath}
where $LL(\theta)$ is the log likelihood (\ref{eq:likelihood}) and $df(\theta)$,  the degree of freedom, is  the number of distinct fixed effects and random effects variances in the selected model. For a given $\Lambda$,  the penalized SAEM algorithm estimates a model ($\theta_{\Lambda}$) with a particular structure: some parameters have the same estimated value (their difference is set to $0$). However, the estimate $\theta_{\Lambda}$ is biased as the penalty shrinks the differences towards $0$. In order to select the optimal structure, it is common practice to compute the BIC with an unbiased version $\tilde{\theta}_{\Lambda}$ of $\theta_{\Lambda}$ that shares the same structure as $\theta_{\Lambda}$. Following the Lars-OLS-Hybrid algorithm \citep{efron2004least}, $\tilde{\theta}_{\Lambda}$ is obtained by  reestimating $\theta$ with a constrained, but unpenalized, SAEM algorithm:
\begin{linenomath}
\begin{gather*}
  \tilde{\theta}_{\Lambda} =  \underset{ \theta } {\operatorname{ArgMin}} (-2LL(\theta) )    \nonumber \\
   \mbox{ s.t }S \footnotesize{\left( \begin{array}{ccc}P\,  \mu \\ P\,  \mbox{diag }\Omega  \end{array} \right)}    = S \footnotesize{\left( \begin{array}{ccc} P\, \hat{\mu}_{\Lambda} \\ P\, \mbox{diag } \hat{ \Omega}_{\Lambda}  \end{array} \right)},  \nonumber
\end{gather*}
\end{linenomath}
where $S(x)$ is the support of vector $x$. This can be seen as a relaxed lasso  \citep{meinshausen2007relaxed} with relaxing parameter set to $0$. The constraint on the support ensures that the solution of the constrained optimization problem has the exact same structure as the solution of the initial penalized estimate. This constrained optimization problem is not trivial. Even if it is an unpenalized problem, it can be solved by the penalized SAEM algorithm with appropriate choices for the adaptive weights: non-null differences are attached to null weights (and are therefore not penalized) while null differences are attached to weights that are high enough to force these differences to be null in the solution $\tilde \theta_\Lambda$. 
Finally, we take:
\begin{linenomath}
\begin{gather*}
  BIC(\Lambda) = -2LL(\tilde{\theta}_{\Lambda}) + \log(N)\times df(\tilde{\theta}_{\Lambda}).  \nonumber
\end{gather*}
\end{linenomath}
This allows the computation of the optimal value  $\Lambda_{BIC}$. Then the final estimator of the procedure is set to $\tilde{\theta}_{\Lambda_{BIC}} $.

\section{Simulated data analysis}
\label{sec:SIM}
Simulations are performed  under  the one compartment model with first order absorption:
\begin{linenomath}
\begin{gather}\label{eq:model1cpt}
f(t,k_{a},Cl,V) = \frac{D k_{a}}{V  k_{a} - Cl} ( e^{-\frac{Cl}{V}t} - e^{-k_{a}t} ), 
\end{gather}
\end{linenomath}
where $k_{a}$ ($h^{-1}$), $Cl$ ($L.h^{-1}$) and $V$ ($L$)  correspond to the absorption rate, the clearance and the volume of distribution, respectively. The administrated dose ($D$) is set to $6$ $mg$.

First, the behavior of the penalized SAEM algorithm is illustrated  on one data set  simulated with 3 groups of subjects.  In particular, regularization paths are presented. Then the impact of high variances on the penalized estimation is studied on  one data set simulated with 3 groups of subjects, and the benefit of adaptive weights introduced in the penalty is shown. Next the influence of the penalty structure on selection performance is studied on $100$ simulated data sets with $5$ groups of subjects. Finally,  we compare our proposal and the standard stepwise forward approach with regard to model selection on $100$ simulated data sets with  $2$ groups of subjects. 
$ $\\

\begin{figure}
\begin{center}
\includegraphics[scale = 0.38]{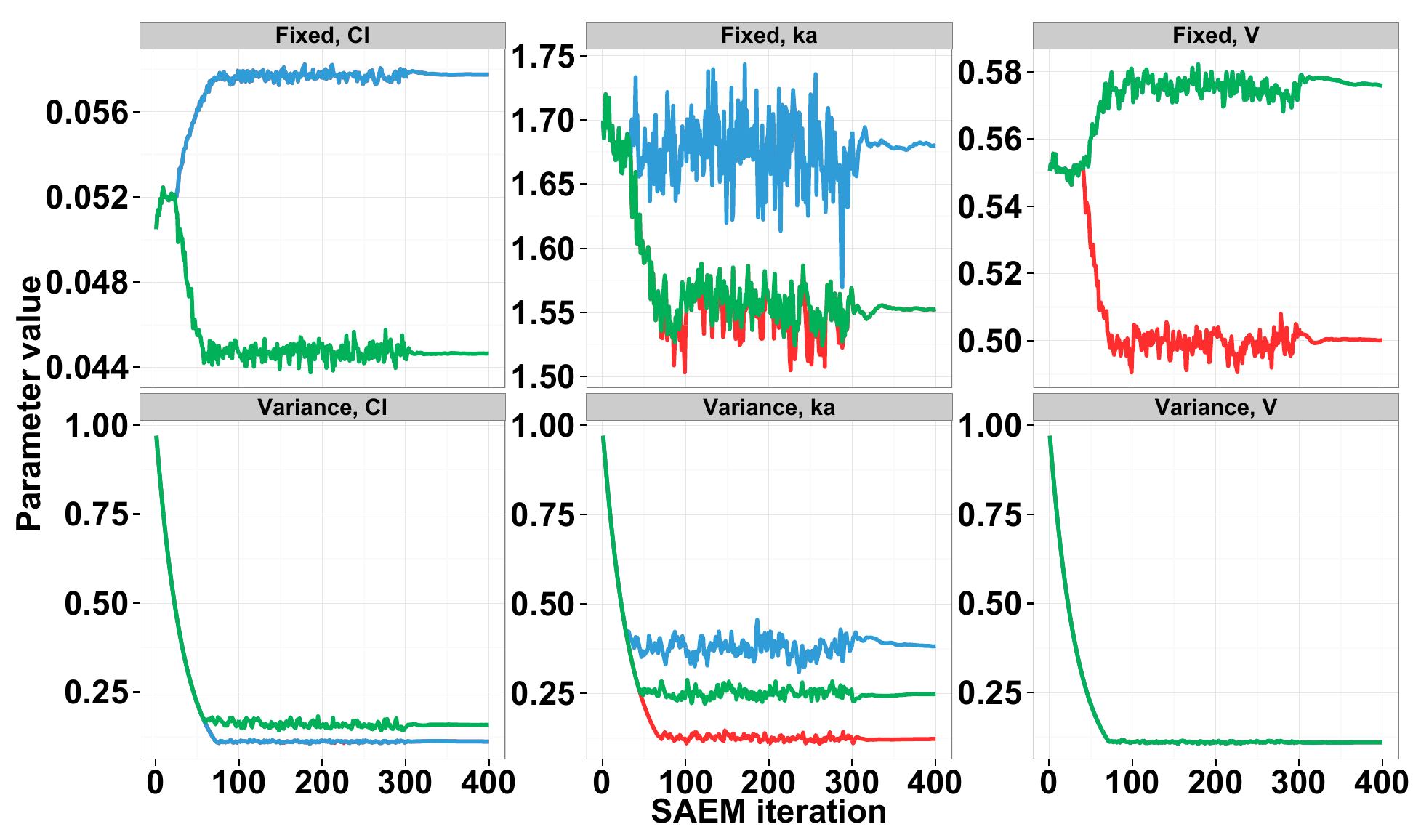}
\caption{Simulated data,   3 groups: evolution of SAEM estimates   with  $\lambda_{F} = 25$ and $\lambda_{V}= 0.013$. Red, blue and green curves correspond  to   estimates of group 1, 2 and 3, respectively.}
\label{fig:EstParam1}
\end{center}
\end{figure}

\begin{figure}
\begin{center}
\includegraphics[scale = 0.38]{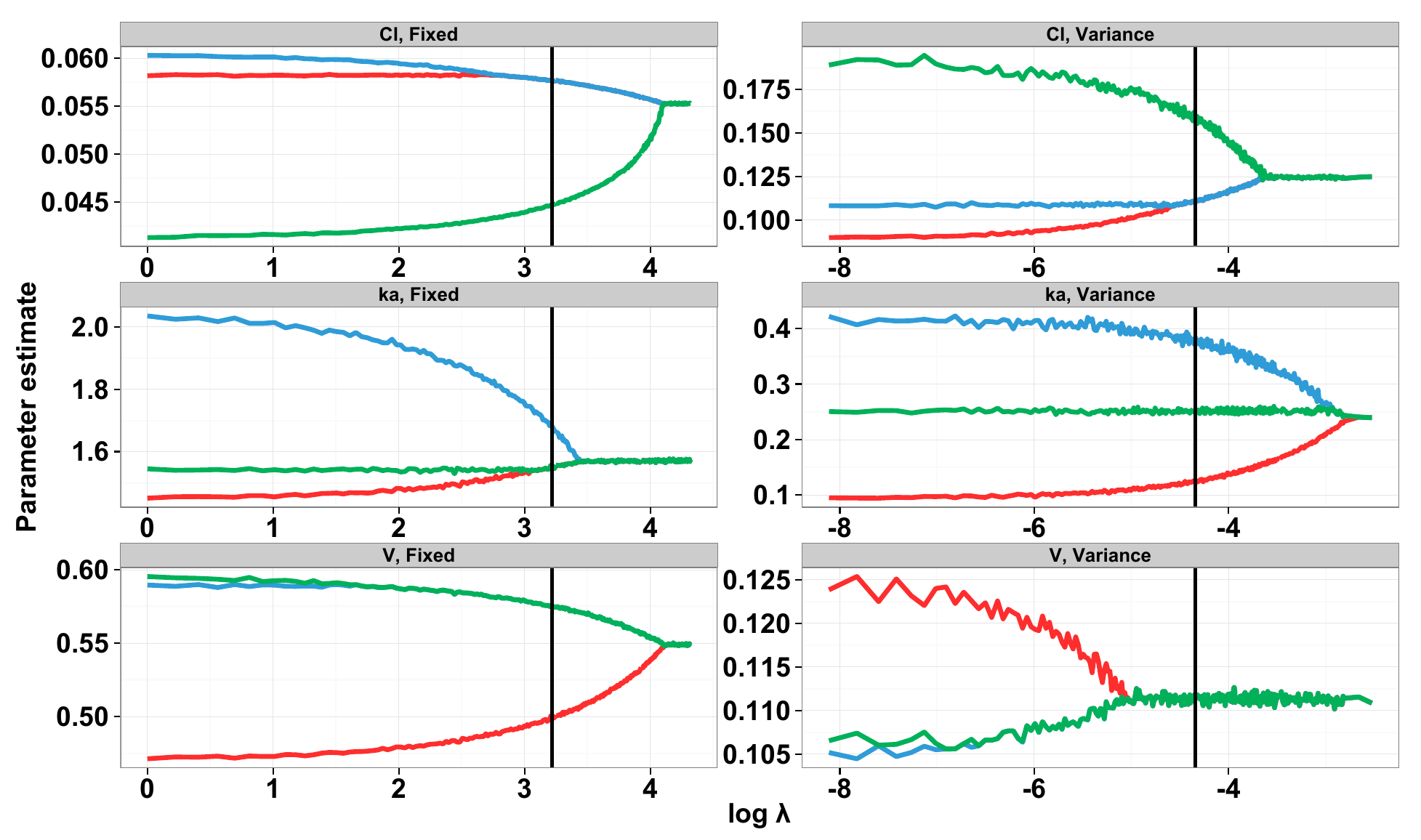}
\caption{Simulated data,   3 groups: regularization paths of SAEM estimates for fixed effects and random effect variances. Red, blue and green curves correspond  to   estimates of group 1, 2 and 3, respectively. Solid black lines corresponds to the lambda values used in Figure  \ref{fig:EstParam1}  $(\lambda_{F} = 25, \lambda_{V}= 0.013)$.}
\label{fig:EstParam2}
\end{center}
\end{figure}

To illustrate the behavior of the penalized SAEM algorithm, one  data set of 3 groups with $N_{g}=100$ subjects per group has been simulated using model (\ref{eq:model1cpt}) and following fixed effects parameters:
\begin{linenomath}
 \begin{eqnarray}
&&\mu_{1,V} =  \mu_{2,V} = 0.48 \mbox{ and } \mu_{3,V} = 0.58.   \nonumber \\
&& \mu_{1,Cl} =  \mu_{2,Cl} = 0.06 \mbox{ and } \mu_{3,Cl} = 0.042.   \nonumber \\
&&\mu_{1,k_{a}} =  \mu_{3,k_{a}} = 1.47 \mbox{ and } \mu_{2,k_{a}} = 2.18.  \nonumber
\end{eqnarray}
\end{linenomath}
 Random effects variances are set to:
 \begin{linenomath}
 \begin{eqnarray}
&& \omega_{1,V}^{2} = \omega_{2,V}^{2}= \omega_{3,V}^{2}= 0.1.  \nonumber \\
&& \omega_{1,Cl}^{2}= \omega_{2,Cl}^{2}= 0.1 \mbox{ and } \omega_{3,Cl}^{2}= 0.2.  \nonumber \\
&& \omega_{1,k_{a}}^{2}= 0.1 \mbox{, }  \omega_{2,k_{a}}^{2}= 0.3 \mbox{ and } \omega_{3,k_{a}}^{2}= 0.2.  \nonumber 
 \end{eqnarray}
 \end{linenomath}
Individual parameters are   log-normally distributed ($h(\phi) = \log(\phi)$). Error model parameters are set to $a=0$ and $b=0.1$. The  penalized  SAEM algorithm is implemented with 400 iterations, { with a clique graph for the fused penalty. In this example, the number of iterations has been chosen so that convergence is clearly attained for all the model parameters. During the first 300 iterations, we use a constant step size equal to 1. Then, during the last 100 iterations, the stochastic approximation scheme is implemented with a step size equal to $\frac{1}{k - 300}$ at iteration $k$. The evolution of each SAEM parameter estimate is plotted along iterations in Figure \ref{fig:EstParam1} for $\lambda_{F} = 37$ and $\lambda_{V} = 0.015$.} For these values of $\lambda_{F}$ and $\lambda_{V}$, the model selected by the algorithm corresponds to the simulated one. Figure \ref{fig:EstParam2}  {presents} the regularization paths of the estimates for both fixed effects and variances of random effects parameters. When increasing $\lambda_{F}$ (or $\lambda_{V}$) values, differences between estimates get smaller and smaller until being null. The number of null differences increases with the value of $\lambda$.

As  {mentioned in the algorithm description, the proximal operation of the variances penalty needs to be numerically approximated when $G>2$. Therefore, computational time is expected to increase} when variances are penalized. Table \ref{table:simu1} presents  {computational times for one run of the penalized SAEM algorithm, for various numbers of groups, with} variances penalized or not. Computational time   {naturally depends on} $\lambda_{F}$ and $\lambda_{V}$; Table  \ref{table:simu1} corresponds to a worst-case scenario (small values of $\lambda_{F}$ and $\lambda_{V}$). 

\begin{table}
\begin{center}
\begin{tabular}{cccc}
\cline{2-4}
                               &  \multicolumn{1}{c}{$G=2$} & $G=3$  &  $G=5$              \\
\hline
\multicolumn{1}{l}{Fixed} & $23$ s & $24$ s & $99$ s      \\
\hline
\multicolumn{1}{l}{Fixed + Variances} & $32$ s & $210$ s & $411$ s  \\
\hline
\hline
\end{tabular}
\end{center}
\caption{Simulated data,  2,  3 or 5 groups: computational time of one run of the penalized SAEM algorithm (400 iterations) with small tuning parameter values on a simulated data set of $N_{g}=100$ subjects per group ($G=2$, $3$ or $5$) with or without penalty on  {the} variance parameters. A clique graph is used for the penalty.}\label{table:simu1}
\end{table}

As discussed in Section \ref{sec:PEN},  {the penalty based on the concentration matrix is not equivalent to that based on the variances}. It could favor differences from parameters with a high variance and then select  {inappropriate} models. This  can be attenuated by rescaling the variances with adaptive weights. We propose the following weighting strategy:
\begin{eqnarray}
&& P_{V}(\Omega_{1}^{-1},\ldots , \Omega_{G}^{-1}) =  \sum_{(g_{1},g_{2})\in \mathcal{E}} \nu_{i}  \sum_{i=1}^{p}  \vert \Omega_{g_{1},ii}^{-1} -\Omega_{g_{2},ii}^{-1} \vert \mbox{ and }  \nu_{i} = \sqrt{\sum_{g=1}^{G} \tilde{\Omega}_{g,ii}}, \nonumber 
\end{eqnarray}
where $\tilde{\Omega}_{g}$  {stands for} the unpenalized estimation of $\Omega_{g}$. To illustrate  this  {approach},  a data set of 3 groups (100 subjects per group) is simulated  {under} model (\ref{eq:model1cpt}) with larger $\omega_V$ and smaller $\omega_{k_{a}}$:

\begin{eqnarray}
 &&\omega_{1,V}^{2} = \omega_{2,V}^{2} = \omega_{3,V}^{2} = 0.3.  \nonumber \\
 &&\omega_{1,Cl}^{2} =  \omega_{2,Cl}^{2} = 0.1,  \mbox{ and } \omega_{3,Cl}^{2} = 0.2.   \nonumber \\
 &&\omega_{1,k_{a}}^{2} = 0.03\mbox{, } \omega_{2,k_{a}}^{2} = 0.075 \mbox{ and } \omega_{2,k_{a}}^{2}= 0.06.   \nonumber 
\end{eqnarray}

\begin{figure}
\begin{center}
\includegraphics[scale = 0.38]{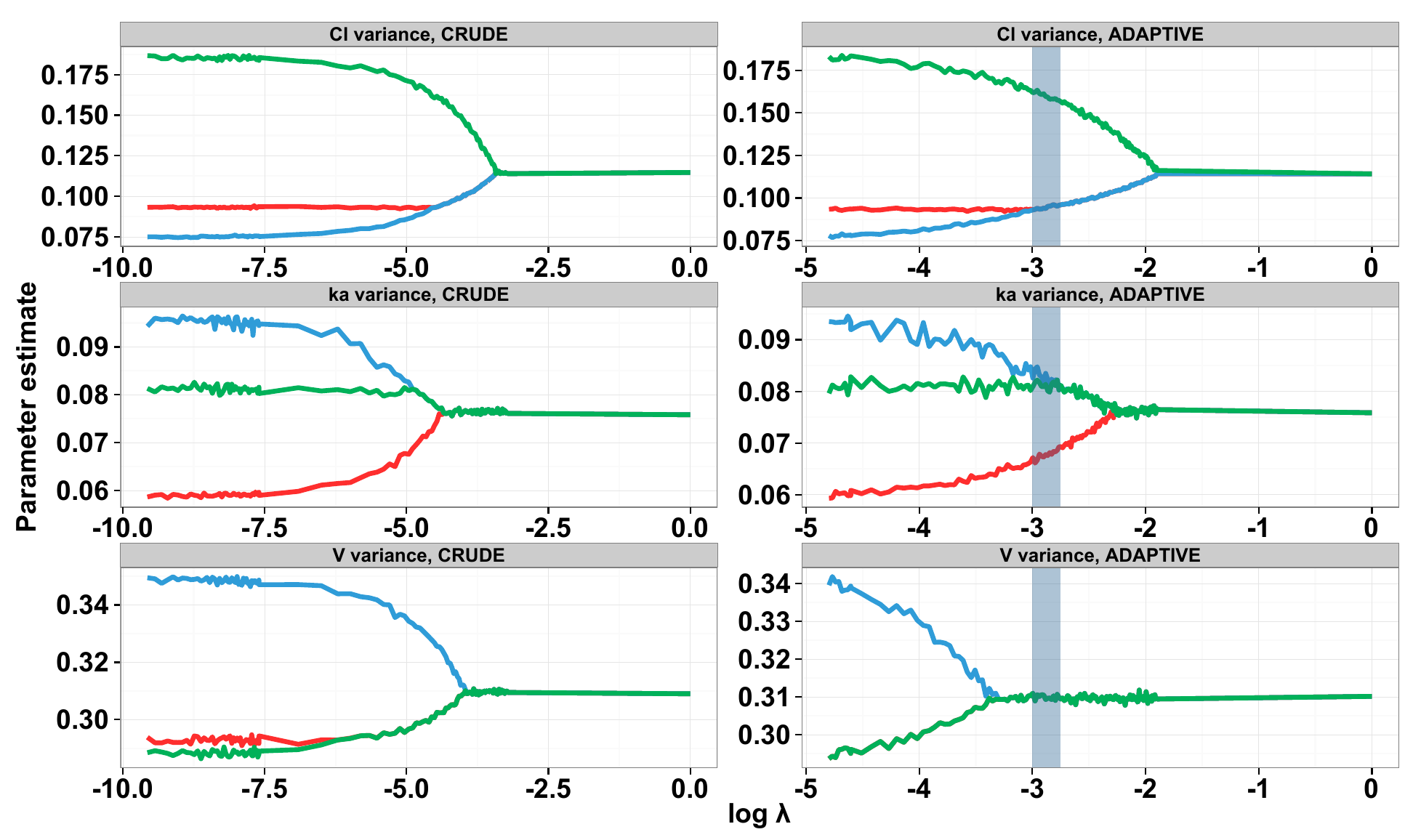}
\caption{Simulated data,    3 groups, large $\omega_V^2$, small $\omega_{k_{a}}^2$: regularization paths of SAEM estimates for random effect variances with (ADAPTIVE) or without (CRUDE) adaptive weights. Red, blue and green curves correspond to the estimates in group $1$, $2$ and $3${, respectively}. Blue shaded areas correspond to  {values  $\Lambda$ returning the simulated model} .}
\label{fig:VAR_SCALING}
\end{center}
\end{figure}

Figure \ref{fig:VAR_SCALING} presents the regularization path of estimates for $\omega_{g,k_{a}}^{2}, \omega_{g,V}^{2}, \omega_{g,Cl}^{2}$ using a clique{-graph in the fused penalty}. Because the $\omega_{g,V}^{2}$ terms are all equal, {estimates of these terms are expected to be fused before the $\omega_{g,k_{a}}^{2}$ and $\omega_{g,Cl}^{2}$ terms}. This is not the case without adaptive weights, and as a consequence, the {simulated} model is not spanned in the regularization path. Adaptive weights correct for this defect and the {simulated} model is spanned by the regularization path (blue shaded areas in Figure \ref{fig:VAR_SCALING} ).

$ $\\

Next, we study the selection of fixed effects differences between groups on simulated data sets{. In particular, the impact of the penalty structure (that is the graph used in the fused penalty) on the proportion of correctly selected models is evaluated on 100 datasets simulated  model (\ref{eq:model1cpt}) with $5$ groups of subjects and with $N_{g}=20$ or $N_{g}=100$.} Fixed effects parameters are set to:
\begin{linenomath}
\begin{eqnarray}
&& \mu_{1,V} = 0.48\mbox{, } \mu_{2,V} = \mu_{3,V} =0.72 \mbox{ and } \mu_{4,V} =\mu_{5,V} =0.96. \nonumber \\
&& \mu_{1,Cl} = \mu_{2,Cl} = 0.06\mbox{, }  \mu_{3,Cl} = \mu_{4,Cl} = 0.03 \mbox{ and } \mu_{5,Cl} =0.015. \nonumber \\
&& \mu_{1,k_{a}} =  \mu_{2,k_{a}} = \mu_{3,k_{a}} = \mu_{4,k_{a}} =\mu_{5,k_{a}} = 1.47. \nonumber 
\end{eqnarray}
\end{linenomath}
{Random effects variances for all the parameters are set to $0.1$}. Individual parameters are log-normally distributed. Error model parameters are set to $a=0$ and $b=0.1$. For each data set, a model is selected using the fused lasso approach on a grid of $100$ $\lambda_{F}$ values with $4$ different penalty structures:
\begin{itemize}
 \item $CH_A, CH$: chain graph with adaptive weights or not.
 \item $CL_A, CL$: clique graph with adaptive weights or not.
 \item $S_{1,A}, S_{1}$: star graph with  {reference set to group 1, } with adaptive weights or not.
 \item $S_{3}$: star graph with  {reference set to group 3, without adaptive weights.}
 \end{itemize} 
Note that the optimal graph  {is the one with vertices exactly corresponding to} the null differences that appear in the simulated model. Thus none of these graphs is optimal for all the parameters. The optimal structure for parameter $\mu_{k_{a}}$ is a clique structure because its value is the same for all the groups. The most appropriate structure for  parameters $\mu_{Cl}$ and $\mu_{V}$ is the chain structure that penalizes all theoretically null differences (unlike  {the} star graph) and  {fewer non-null} differences than the clique graph.  {To recap, there is a hope to select the true model only when using chain or clique graphs (and not when using star graphs)}.   {As previously suggested by \citep{viallon2014robustness}, adaptive weights generally makes fused lasso estimates more robust to graph misspecification. To confirm this result in our context the penalized SAEM algorithm is implemented with the $4$ mentioned graphs with and without adaptive weights.} 
 The performance of each penalty structure is evaluated  by comparing the selected model ($P \tilde \mu_{s}$) to the true model ($P \mu$) for each parameter, on each simulated data set ($s=1,...,100$){, with $\tilde \mu_{s}$ the final estimate obtained by the fused lasso procedure and $P$ a matrix that encodes the differences under study}. For the $CH$ and  $CL$ penalties, $P$ is the $(G-1)\times G$ matrix defined  by the chain graph ($P=P$ with $P_{i,i}=1$, $P_{i,i+1}=-1$ and $0$ elsewhere). For the $S_{1}$ and  $S_{3}$ penalties, the matrix $P$ is the one defined  by their respective star graph. When considering the whole fixed effect model, the number of correctly selected model is:
  \begin{linenomath}
  \begin{eqnarray}
  \frac{1}{100}\sum_{s=1}^{100} \mathbbm{1}_{P \tilde \mu_{V,s} = P \mu_{V}} \times \mathbbm{1}_{P \tilde \mu_{Cl,s} = P \mu_{Cl}}  \times  \mathbbm{1}_{P \tilde \mu_{k_{a},s} = P \mu_{k_{a}}}. \nonumber
  \end{eqnarray}
  \end{linenomath}

   \begin{table}
\begin{center}
\renewcommand{\arraystretch}{1.5}
\begin{tabular}{*{5}{c}}
\cline{2-5}
\multicolumn{1}{l}{}  & \multicolumn{4}{c}{$N_{g}=20$}  \\
\cline{2-5}
\multicolumn{1}{l}{}  & $CH$ & $CH_{A}$& $CL$ & $CL_{A}$     \\
\hline
 \multicolumn{1}{l}{Whole} & $15\mbox{ }(8-22)$&$39\mbox{ }(29-48)$&$8\mbox{ }(3-13)$ &  $32\mbox{ }(23-41)$    \\
\cline{2-5}
 \multicolumn{1}{l}{$\mu_V$} & $53\mbox{ }(43-63)$&$71\mbox{ }(62-80)$&$33\mbox{ }(24-42)$ & $56\mbox{ }(46-66)$  \\
\cline{2-5}
 \multicolumn{1}{l}{$\mu_{Cl}$} & $41\mbox{ }(31-51)$&$86\mbox{ }(79-93)$&$29\mbox{ }(20-38)$ &  $69\mbox{ }(60-78)$\\
\cline{2-5}
 \multicolumn{1}{l}{$\mu_{k_{a}}$} & $61\mbox{ }(51-70)$&$68\mbox{ }(59-73)$&$64\mbox{ }(55-73)$ &  $81\mbox{ }(73-89)$ \\
\hline
\hline

\multicolumn{1}{l}{}  & \multicolumn{4}{c}{$N_{g}=100$} \\
\cline{2-5}
\multicolumn{1}{l}{}   & $CH$& $CH_{A}$& $CL$ & $CL_{A}$    \\
\hline
 \multicolumn{1}{l}{Whole} &   $25\mbox{ }(17-33)$& $59\mbox{ }(49-69)$&$28\mbox{ }(19-37)$ & $55\mbox{ }(45-65)$    \\
\cline{2-5}
 \multicolumn{1}{l}{$\mu_V$} &  $54\mbox{ }(44-64)$& $80\mbox{ }(72-89)$&$52\mbox{ }(42-62)$ &  $81\mbox{ }(73-89)$ \\
\cline{2-5}
 \multicolumn{1}{l}{$\mu_{Cl}$} & $55\mbox{ }(45-65)$&$78\mbox{ }(70-86)$&$54\mbox{ }(44-64)$ &  $70\mbox{ }(61-80)$ \\
\cline{2-5}
 \multicolumn{1}{l}{$\mu_{k_{a}}$} & $77\mbox{ }(69-85)$& $75\mbox{ }(66-83)$&$77\mbox{ }(69-85)$ &  $87\mbox{ }(80-93)$ \\
\hline
\hline

\end{tabular}
\caption{Simulated data,  5 groups:  {Proportion of correctly selected models over 100 simulations  (with $95\%$ confidence interval between brackets) for the whole fixed effects model and fixed effects model restricted to $\mu_V$, $\mu_{Cl}$ or $\mu_{k_{a}}$. Various penalty structures are compared:  chain ($CH$), adaptive chain ($CH_{A}$), clique ($CL$) and adaptive clique ($CL_{A}$). Further keep in mind that model is said to be correctly selected whenever $P \tilde \mu_{s} =P \mu$ here.} }
\label{table:SIM5GRP}
\end{center}
\end{table}

 Table \ref{table:SIM5GRP} shows the results for $CH$ and $CL$. When $N_{g}=20$, the chain graph  {globally performs the best}.  When  {$N_{g}=100$,  the chain and clique graphs perform similarly}. In addition,    adaptive weights improves performance: in particular, the $CL_A$ performs similarly to $CH_A$.  {On these examples, the clique graph appears as } a good candidate when there is no prior information on data structure. This results tends to confirm the asymptotic optimality result of the clique-based strategy with adaptive weights that was obtained for generalized linear models \citep{viallon2014robustness}.

\begin{table}
\begin{center}
\renewcommand{\arraystretch}{1.5}
\begin{tabular}{*{4}{c}}
\cline{2-4}
  \multicolumn{1}{l}{} & \multicolumn{3}{c}{$N_{g}=20$} \\
\cline{2-4}
\multicolumn{1}{l}{}    & $S_{1}$ & $S_{1,A}$&$S_{3}$ \\
\hline
 \multicolumn{1}{l}{Whole}   & $1\mbox{ }(0-3)$ & $6\mbox{ }(1-11)$&$26\mbox{ }(17-34)$   \\
\cline{2-4}
 \multicolumn{1}{l}{$\mu_V$}  & $69\mbox{ }(60-78)$  & $72\mbox{ }(63-81)$&$57\mbox{ }(47-67)$  \\
\cline{2-4}
 \multicolumn{1}{l}{$\mu_{Cl}$}  & $36\mbox{ }(26-45)$ &$87\mbox{ }(80-93)$&$83\mbox{ }(76-90)$ \\
\cline{2-4}
 \multicolumn{1}{l}{$\mu_{k_{a}}$}  &  $12\mbox{ }(6-18)$ & $23\mbox{ }(15-31)$&$58\mbox{ }(48-68)$   \\
\hline
\hline
\multicolumn{1}{l}{} & \multicolumn{3}{c}{$N_{g}=100$} \\
\cline{2-4}
\multicolumn{1}{l}{}    & $S_{1}$  & $S_{1,A}$ & $S_{3}$\\
\hline
 \multicolumn{1}{l}{Whole}   & $8\mbox{ }(3-13)$& $56\mbox{ }(46-66)$&$49\mbox{ }(39-59)$  \\
\cline{2-4}
 \multicolumn{1}{l}{$\mu_V$}  & $100\mbox{ }(99-100)$ & $100\mbox{ }(99-100)$&$90\mbox{ }(84-96)$ \\
\cline{2-4}
 \multicolumn{1}{l}{$\mu_{Cl}$}  & $29\mbox{ }(20-38)$ & $77\mbox{ }(69-85)$&$93\mbox{ }(88-98)$ \\
\cline{2-4}
 \multicolumn{1}{l}{$\mu_{k_{a}}$}  & $28\mbox{ }(19-37)$ & $63\mbox{ }(53-72)$&$60\mbox{ }(50-69)$  \\
\hline
\hline
\end{tabular}
\caption{Simulated data,  5 groups:  Proportion (with $95\%$ confidence interval between brackets) of correctly selected models over 100 simulations when edges under study corresponds to a star graph. Results are given for the whole fixed effects model, fixed effects of $\mu_V$, $\mu_{Cl}$ or $\mu_{k_{a}}$.  Different   penalty structures are considered: star with group $1$ as reference ($S_{1}$), adaptive star with group $1$ as reference ($S_{1,A}$) and star with group $3$ as reference ($S_{3}$). Further keep in mind that model is said to be correctly selected here whenever $P \tilde \mu_{s} =P \mu$.  }
\label{table:SIM5GRPbis}
\end{center}
\end{table}
  
Table \ref{table:SIM5GRPbis} shows the results  {when using star graphs} ($S_{1}$, $S_{1,A}$ and $S_{3}$). Keep in mind that the star structure does not correspond to the real structure of the model: here only differences ``encoded'' in the star graph can be set to zero. Table \ref{table:SIM5GRPbis} highlights the dramatic influence of the reference group on the performance when using star graphs. It is particularly true for   $\mu_{Cl}$. Indeed with $S_1$, theoretical values of $\mu_{2,Cl}$, $\mu_{3,Cl}$, $\mu_{4,Cl}$ and $\mu_{5,Cl}$ are distributed in an unbalanced way around $\mu_{1,Cl}$: $\mu_{3,Cl}$, $\mu_{4,Cl}$ and $\mu_{5,Cl}$ are lower than $\mu_{1,Cl}$. The penalty unexpectedly tends first to fuse $\mu_{1,Cl}$ with $\mu_{3,Cl}$, $\mu_{4,Cl}$ and $\mu_{5,Cl}$. The adaptive version $S_{1,A}$ seems to mitigate this phenomenon when sample size is large ($N_{g}=100$). This behavior is not observed with $S_3$, probably because theoretical parameters value of non reference groups are distributed in a more balanced way around $\mu_{3,Cl}$.
$ $\\


Finally, the joint selection of fixed effects and random effects variances is evaluated through 100 simulated  {data sets} using   model (\ref{eq:model1cpt}) with only two groups for computational time reasons. Individual parameters are log-normally distributed. Error model parameters are set to $a=0.2$ and $b=0.02$. Fixed effects parameters are:
\begin{linenomath}
\begin{eqnarray}
 &&\mu_{1,V} = 0.48 \mbox{ and }  \mu_{2,V} = 0.58.   \nonumber \\
 &&\mu_{1,Cl} =  0.060 \mbox{ and }  \mu_{2,Cl} =  0.042.  \nonumber \\
 &&\mu_{1,k_{a}} =  \mu_{2,k_{a}} = 1.47.   \nonumber
\end{eqnarray}
\end{linenomath}
Random effects variances are:
 \begin{linenomath}
 \begin{eqnarray}
 &&\omega_{1,V}^{2} = \omega_{1,V}^{2} = 0.1. \nonumber \\
 && \omega_{1,Cl}^{2} = 0.1 \mbox{ and } \omega_{2,Cl}^{2} = 0.21. \nonumber \\
 &&\omega_{1,k_{a}}^{2} = 0.1 \mbox{ and } \omega_{2,k_{a}}^{2} = 0.21. \nonumber 
\end{eqnarray} 
\end{linenomath}
For each data set, the best model is selected using BIC based on the penalized SAEM algorithm estimations with a grid of $100$ ($\lambda_{F}$,$\lambda_{V}$) values. For comparison purpose, the selection approach based on a BIC  forward stepwise method is also implemented using the constrained SAEM algorithm ({see Section \ref{sec:BIC}}). This stepwise method includes $2$ steps: i) assuming the variances of random effects to be different between the groups,  the fixed effect model is selected by BIC comparison, ii) using the selected  fixed effects model,  the variance model is selected by BIC comparison. The performance of the two methods is evaluated  by comparing the selected model to the true model. The selection of an optimal model took approximately $10$ min with the stepwise strategy and $56$ min with the penalized approach ({these computational times correspond to averages over the $100$ data sets with $N_{g}=100$}). Table \ref{tab:simu2gpesPct} presents the proportion of correctly selected models for the fixed effects model, the variances model and the whole model. On this synthetic example, our approach have significantly better selection performance for the variance model. Both methods gives similar results for the fixed effects model selection. Supplementary  {Table $1$ further shows that models returned by our approach also tend to be too complex for small sample sizes. In particular, $\mu_{k_{a}}$ and $\omega_{V}^{2}$ are theoretically equal in the $2$ groups,  but the fused lasso returns a non-null difference  {between} these two parameters more often than the stepwise approach}. 
\begin{table}
\begin{center}
\begin{tabular}{ccccccccccccc}
\cline{2-4}
			  & \multicolumn{3}{c}{Fixed effects model}   \\
\hline
             \multicolumn{1}{c}{$N_{g}$}  &  \multicolumn{1}{c}{$25$} & $50$  & $100$             \\
\hline\hline
\multicolumn{1}{l}{Stepwise Forward} & $37\mbox{ }(27-46)$ &$71\mbox{ }(62-80)$ & $63\mbox{ }(54-72)$     \\
\hline
\multicolumn{1}{l}{Fused LASSO} & $44\mbox{ }(34-54)$ & $66\mbox{ }(57-75)$ & $69\mbox{ }(60-78)$  \\
\hline
\hline \\
\cline{2-4}
   & \multicolumn{3}{c}{Variances model}   \\
\hline
             \multicolumn{1}{c}{$N_{g}$}  &  \multicolumn{1}{c}{$25$} & $50$  & $100$              \\
\hline\hline
\multicolumn{1}{l}{Stepwise Forward} & $13\mbox{ }(6-19)$ & $30\mbox{ }(21-39)$ & $49\mbox{ }(39-59)$     \\
\hline
\multicolumn{1}{l}{Fused LASSO} & $40\mbox{ }(30-50)$ & $50\mbox{ }(40-60)$ & $75\mbox{ }(67-83)$ \\
\hline
\hline \\
\cline{2-4}
			  & \multicolumn{3}{c}{Whole model}  \\
\hline
             \multicolumn{1}{c}{$N_{g}$}  &  \multicolumn{1}{c}{$25$} & $50$  & $100$            \\
\hline\hline
\multicolumn{1}{l}{Stepwise Forward} & $8\mbox{ }(3-13)$ &$20\mbox{ }(12-28)$ & $33\mbox{ }(24-42)$      \\
\hline
\multicolumn{1}{l}{Fused LASSO} & $14\mbox{ }(7-21)$ & $33\mbox{ }(24-42)$ & $43\mbox{ }(33-53)$ \\
\hline
\hline
\end{tabular}
\end{center}
\caption{Simulated data,   2 groups:  {proportion} of correctly selected models on  100 simulated datasets for the fixed effects model, the variances model and the whole model. Results are given for the fused lasso and the stepwise forward approaches.}\label{tab:simu2gpesPct}
\end{table}

\section{Real data analysis}
\label{sec:CROSS}
We now illustrate our approach on a real data example. Dabigatran etexilate ($DE$) is an oral anticoagulant drug used for the prevention of venous thromboembolism after orthopedic surgery and stroke in patients with atrial fibrillation. Its has a low bioavailability (fraction of administrated dose that reaches the systemic circulation){, typically} below  $7\%$. It is mainly due to a solubility problem and to the P-glycoprotein (P-gp) efflux that has an {``anti-absorption"} function.  P-gp inhibitors can increase Dabigatran bioavailability by improving its absorption \citep{delavenne2013semi}. {But} the addition of P-gp inhibitors could also lead to overdosing and adverse event like hemorrhage. 

\begin{figure}
\begin{center}
\includegraphics[scale = 0.35]{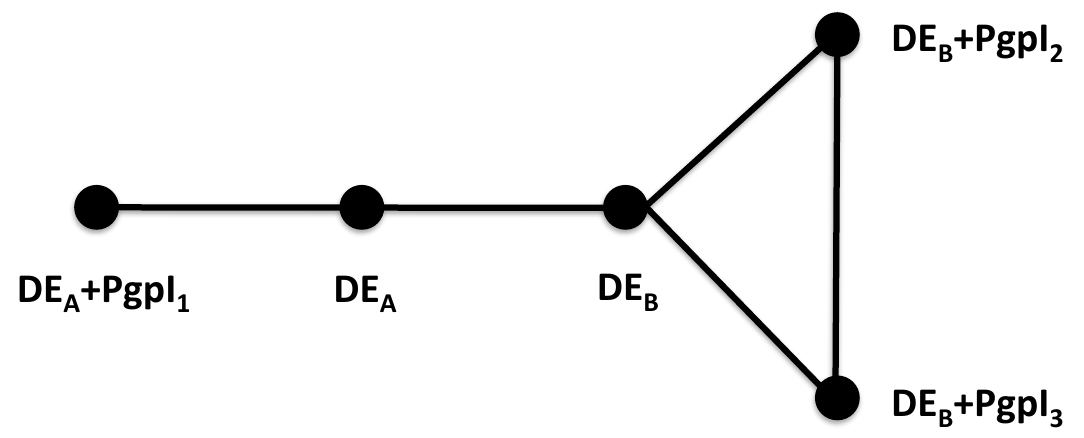}
\caption{Graph used for the penalty of DE pooled data.} 
\label{fig:GraphAppli}
\end{center}
\end{figure}

Data from two crossover clinical trials are considered. The two studies were conducted with two different dosing regimens for $DE$. The first trial, a two way crossover trial with $10$ subjects, evaluates the interaction  between $DE$  (dosing regimen A) and P-Gp inhibithor $1$ ($PgpI_{1}$). The second trial, an incomplete three way crossover trial with $9$ subjects, evaluates the interaction  between $DE$  (dosing regimen B), P-Gp inhibithor $2$ ($PgpI_{2}$) and P-Gp inhibithor $3$ ($PgpI_{3}$). Data from the two trials are pooled and five groups of subjects are defined:
\begin{itemize}
\item $DE_{A}$: DE with dosing regimen A alone ($10$ subjects).
\item $DE_{A}+PgpI_{1}$: DE with dosing regimen A alone plus P-Gp inhibithor $1$ ($10$ subjects).
\item $DE_{B}$: DE with dosing regimen B ($DE_{B}$) alone ($9$ subjects).
\item $DE_{B}+PgpI_{2}$: DE with dosing regimen B alone plus P-Gp inhibithor $2$ ($9$ subjects).
\item $DE_{B}+PgpI_{3}$: DE with dosing regimen B alone plus P-Gp inhibithor $3$ ($9$ subjects).
\end{itemize}
In each group, dabigatran blood concentration {pharmacokinetics})  is measured for each patient at 10 sampling times after oral drug administration. The following pharmacokinetic model with one compartment and an inverse {G}aussian absorption is used:

\begin{linenomath}
\begin{eqnarray}
\begin{cases} \frac{dA_{c}}{dt} = IG(t) - \frac{Cl}{V_{c}}A_{c} \\ IG(t) = Dose \times F \times \sqrt{\frac{MAT}{2\pi CV^{2} t^{3}}} \times e^{\frac{-(t-MAT)^{2}}{2CV^{2}MATt}} \end{cases},  \nonumber
\end{eqnarray}
\end{linenomath}
where $A_{c}$ corresponds to the amount of dabigatran {present} in the blood. The absorption parameters $F$, $MAT$ and $CV$ correspond to bioavailability, mean absorption time and coefficient of variation of the absorption rate respectively. Finally parameters $Cl$ and $V_{c}$  are the elimination clearance and the volume of {the } central compartment. Individual parameters are supposed {to be } log-normally distributed ($h(\phi) = \log(\phi)$).

Estimating the bioavailability with only data from orally administrated drug is an ill-posed problem. Indeed, a decreased value for $F$ could be balanced by smaller $V$ and $Cl$ values.  In order to regularize this problem, we add prior distributions on both $V$ and $Cl$ fixed parameters  \citep{weiss2012modeling} based on previously published results \citep{blech2008metabolism}. In this case, fixed parameters update is done by solving the following optimization problem:

\begin{linenomath}
\begin{gather}
\left( \mu^{1}_{k+1} ,\ldots, \mu^{G}_{k+1}  \right) =  \underset{ \mu } {\operatorname{ArgMax}} \mbox{ }   \sum_{g=1}^{G}  \tilde{Q}_{k}(\mu^{g},\Omega_{k}^{g},a_{k},b_{k}) -\frac{1}{2} \sum_{g=1}^{G}(\mu^{g} - \mu^{g}_{\star})^{t}V_{\star}^{g^{-1}}(\mu^{g} - \mu^{g}_{\star})    -  \lambda_{F} \Vert P \mu  \Vert_{1}, \nonumber
\end{gather}
\end{linenomath}
with {a Gaussian prior distribution $\mathcal{N}(\mu^{g}_{\star},V^{g}_{\star})$ for $\mu^{g}$}. 
Due to the small number of subjects per group, only differences between groups for the bioavailability parameter $F$ are analyzed. The penalized SAEM algorithm is applied to this model penalizing fixed effect and random effects variance of bioavailability ($F$). Parameters $V_{c}$, $Cl$, $MAT$and $CV$ are supposed equal between the   groups. High values for the adaptive weights were used for the corresponding differences to ensure they are null across groups. This assumption seems reasonable as: i) subjects are highly comparable due to very stringent inclusion criterions and ii) P-Gp inhibitors do not seem to influence $MAT$ and $CV$ \citep{delavenne2013semi,ollier2015vitro}. The penalized SAEM algorithm is applied using the graph structure depicted in Figure \ref{fig:GraphAppli} and a grid composed of $400$ pairs of $\lambda_{F}$ and $\lambda_{V}$ values. 

\begin{figure}
\begin{center}
\includegraphics[scale = 0.35]{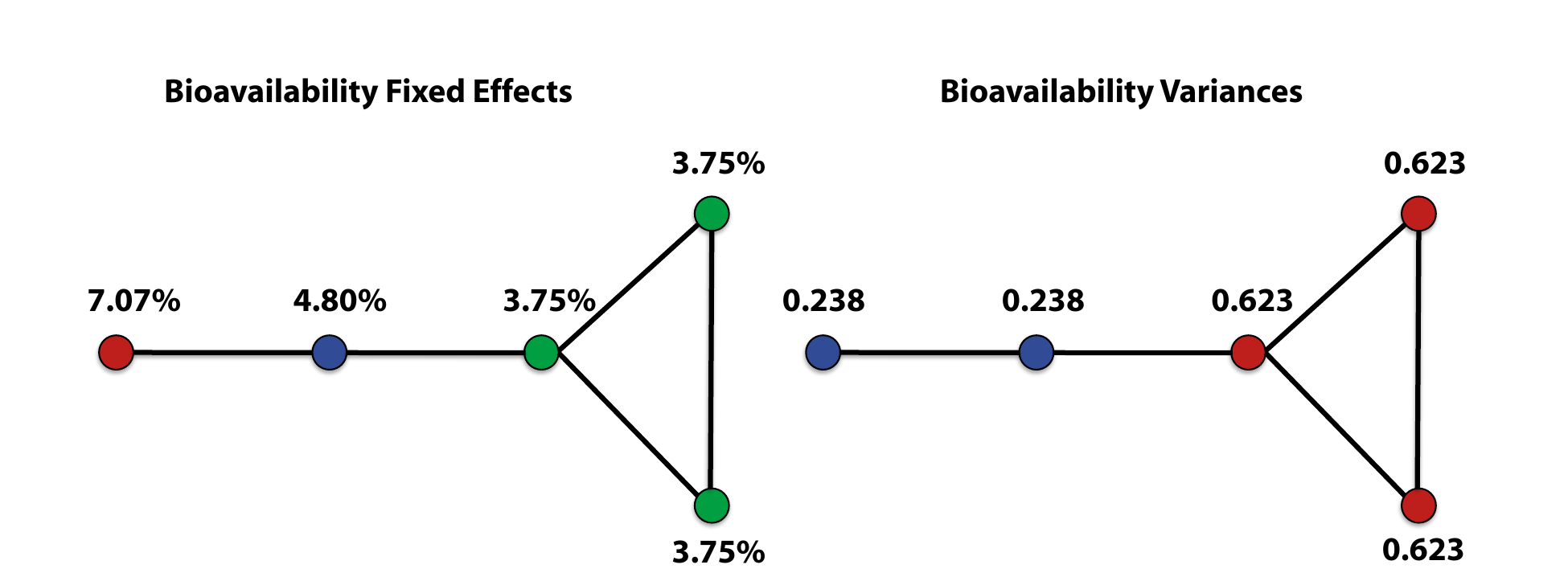}
\caption{Model selected by the BIC and unpenalized re-estimation of the bioavailability parameters from the real data. Groups with  {identical} color share equal estimates.} 
\label{fig:GraphAppli_RES}
\end{center}
\end{figure}

\begin{figure}
\begin{center}
\includegraphics[scale = 0.3]{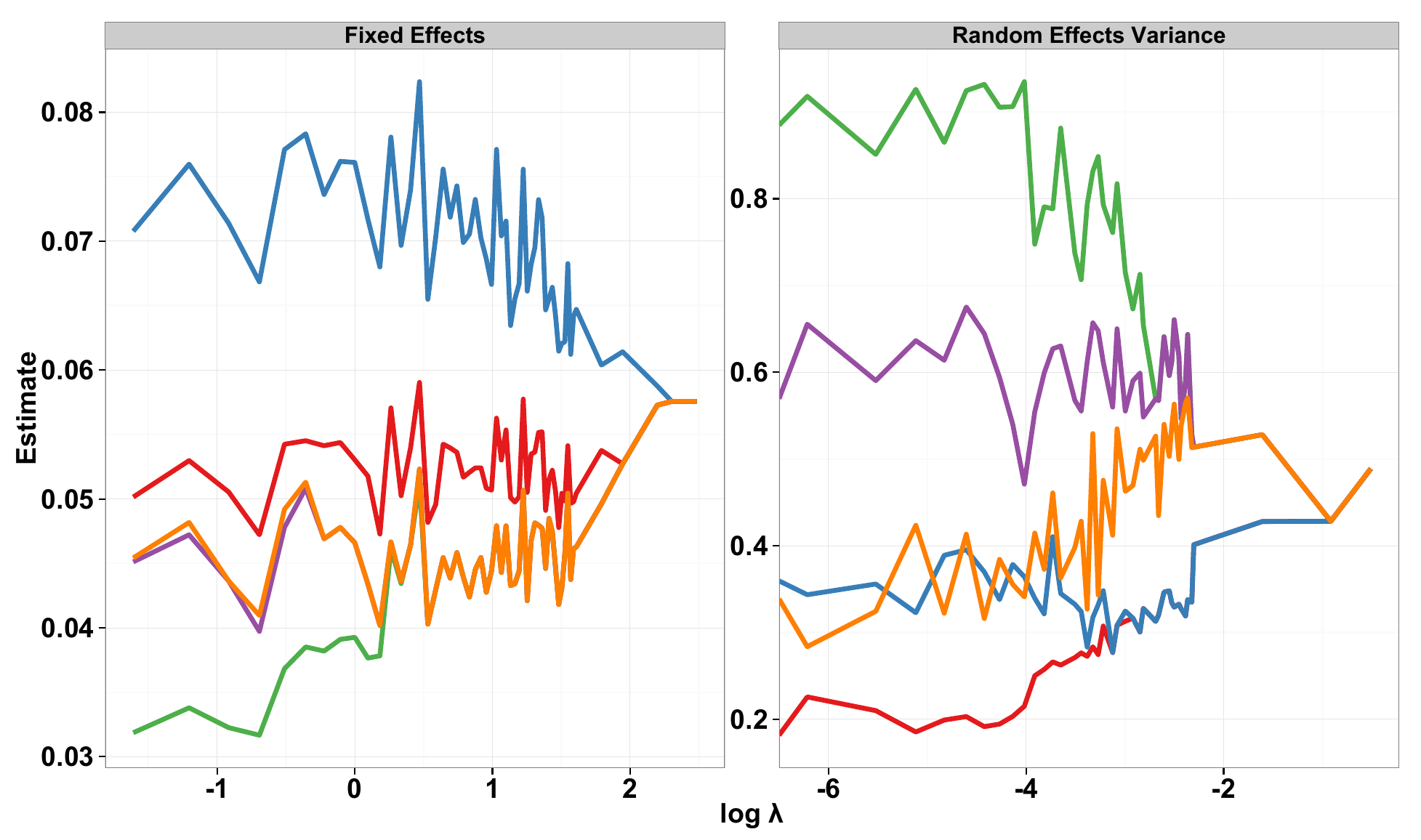}
\caption{Regularization path for both fixed and variance bioavailability parameters from the pooled DE data set. Red, bleu, green, purple and orange  lines correspond to $DE_{A}$,  $DE_{A}+PgpI_{1}$, $DE_{B}$,  $DE_{B}+PgpI_{2}$ and $DE_{B}+PgpI_{3}$ respectively.}
\label{fig:REGPATHAPPLI}
\end{center}
\end{figure}

The optimal model selected by the $BIC$ is shown in Figure \ref{fig:GraphAppli_RES}.  {Regarding} fixed effects, the bioavailability is different between the two dosing regimens. It is  {probably} the consequence of the very low and pH-dependant solubility of DE. As the dosing regimen B was the lowest, then the smaller the dose, the lower the DE  {solubility}. Among the three P-Gp inhibitors, only $PgpI_{1}$ is associated to an increase  {of} DE bioavailability. It is not surprising since $PgpI_{1}$ is known to be a strong P-Gp inhibitor. $PgpI_{2}$ and $PgpI_{3}$  inhibit P-Gp much less in in-vitro experiment. Concerning random effects variances, a higher variance is estimated for dosing regimen B, which again is certainly  {related} to solubility. Finally, Figure \ref{fig:REGPATHAPPLI} shows the regularization path of both fixed effects and variances.

\section{Discussion}

In this paper, we present a fused lasso penalized version of the  SAEM algorithm. It allows the introduction of sparsity in the difference between group parameters for both fixed effects and variances of random effects. This algorithm is designed to iteratively maximize the penalized  conditional expectation of the complete data likelihood. Simulation results show that this algorithm has good empirical convergence properties. The theoretical study of this algorithm will be the scope of future work. The penalized approach was compared to a stepwise forward algorithm.  {This stepwise approach is faster to compute in the case of $G=2$ groups, but this difference in computational time will tend to be smaller as $G$ gets bigger because of the exponential growth of the number of models to be compared.  Moreover, more parsimonious grid could be constructed for the regularization parameters values, which would accelerate our approach. Finally, stepwise forward approaches  are known to generally suffer from high variability, for instance under generalized linear models \citep{oelker2014regularization}.}

Several extensions  {of our work} could be proposed. First,  the  {assumption that the  covariance matrix} is diagonal  might be too strong. For example, in pharmacokinetics the clearance $Cl$ and the volume of distribution parameter may be strongly correlated.  Neglecting this correlation could have important consequences on the model prediction properties. Moreover,  the penalty used in this work does not allow  {the selection of random effects}. One way to tackle these two issues would be to directly penalize the {covariance} matrix (instead of its inverse), which could be achieved by using the parametrization described by  \cite{bondell2010joint}.  {In addition, and as mentioned by one of the reviewers, the methodology described in this paper can be used under mixed effects models (linear or nonlinear) not only for continuous data but also for count, categorical or survival data as long as it keeps the following hierarchical structure:}
\begin{linenomath}
\begin{gather}
p(y_{g},\phi_{g};\theta_{g}) = p(y_{g} \vert \phi_{g};\theta_{g})p( \phi_{g} \vert \theta_{g}). \nonumber
\end{gather}
\end{linenomath}
Moreover, residual error parameters were considered to be independent from the group structure but,  they could be estimated within each group and then being penalized. In this work, group sizes are supposed equal or not too different, which is often the case in pharmacokinetic. The algorithm could be easily modified by introducing the group size in the sum of the  group conditional expectation   \citep{danaher2011joint}:
\begin{linenomath}
  \begin{gather*}
 \sum_{g=1}^{G} N_{g}\tilde{Q}_{k}(\mu_{g},\beta_g,\Omega_{g,k},a_{k},b_{k}).
 \end{gather*} 
 \end{linenomath}
Concerning the selection of tuning parameters,  {criteria other} than BIC have been used for generalized linear models. The cross-validated prediction error may be particularly useful especially for high dimensional data since the unpenalized re-estimation of the log-likelihood can not always be done. For NLME, this criterion has already been studied by  \cite{colby2013cross} and could be easily implemented. Finally a last improvement, subject of a future work, is the extension to NLMEMs including more than one level of random effects \citep{panhard2009extension}. Indeed in this paper the method is applied to data from a cross-over trial, where each subject receives the two treatment modalities. This information was neglected and  the five groups were considered as independent which could lead to spurious association when inter occasion variability is high.

\bibliographystyle{abbrvnat}
\bibliography{BIB}

\end{document}